\documentclass[12pt]{article}

\usepackage[dvips]{graphicx}
\usepackage[english]{babel}

\usepackage{amssymb,epsfig}
\usepackage{amsmath}

\usepackage{slashed}
\usepackage[active]{srcltx}

\newcommand{\ms}{\mskip 1.5mu}

\begin{document}

\begin{center}
\textbf{\LARGE  Proton dissociation into three jets}

\vspace*{1.6cm}

{\large V.M. Braun$\ms{}^1$, D.Yu.\ Ivanov$\ms{}^{2}$ and A.
Sch\"afer$\ms{}^1$ }

\vspace*{0.4cm}

\textsl{%
$^1$ Institut f\"ur Theoretische Physik, Universit\"at Regensburg,
D-93040 Regensburg, Germany \\
$^2$ Sobolev Institute of
Mathematics, 630090 Novosibirsk, Russia }

\vspace*{0.8cm}

\textbf{Abstract}\\[10pt]
\parbox[t]{0.9\textwidth}{
We explore the possibility to observe hard exclusive three-jet production
in early LHC runs, corresponding to diffractive
dissociation of the incident proton into three jets with large but compensating transverse
momenta.  This process is
sensitive to the proton unintegrated gluon distribution at small $x$
and to the distribution of the three valence quarks in the proton
at small transverse distances. The corresponding cross section is
calculated using an approach based on $k_t$ factorization.
According to our estimates, observation of hard diffractive three-jet production  at LHC
is feasible for jet transverse momenta $q_\perp \sim 5$~GeV.}
\end{center}

\vspace*{1cm}



{\large \bf 1.}~~ The physics potential of forward detectors at LHC, within and beyond the standard model,
is attracting a lot of attention,
cf. \cite{Joram:2006kf,Royon:2007ah,Rouby:2008aj,Albrow:2008pn,d'Enterria:2008is}.
In this Letter we explore the possibility to observe hard exclusive diffractive
dissociation of a proton into three hard jets in proton-proton collisions
\begin{equation}
p(p_1)+p(p_2)\to jet(q_1)+jet(q_2)+jet(q_3)+ p(p_2^\prime)\,,
\label{process}
\end{equation}
cf. \cite{Frankfurt:2002jq}
In this process one proton stays intact and the other one dissociates into a system of three hard jets
separated  by a large rapidity gap from the recoil proton, see Fig.~1.
The main aim of our study is to estimate the cross section of this reaction  and  the corresponding event rates at LHC
and Tevatron.

Note that  we are interested in {\it exclusive}\  three--jet production which constitutes a small fraction of the inclusive
single diffraction cross section.
The exclusive and inclusive mechanisms have  different final state topologies and can be distinguished experimentally.
A characteristic quantity is e.g. the ratio $R_{jets}$ of the three-jet mass to the total invariant mass of the system
produced in the diffractive interaction.
Exclusive production corresponds to  the region where  $R_{jets}$ is close to unity.
This strategy was used recently at the Tevatron  \cite{Aaltonen:2007hs}
where  central exclusive dijet production, $p\bar p\to p +jet+jet+\bar p$, in double--Pomeron collisions
was measured for the first time.

Exclusive dijet production in the central region has much in common with
the exclusive Higgs boson production process, $p\bar p\to p +H+\bar p$.
In \cite{Khoze:2008cx} it was argued that studies of exclusive dijet production and other diffractive processes
at the early data runs of the LHC can provide valuable checks of the different components of the formalism.
Indeed, this was the main motivation for Tevatron experiment.
The exclusive 3-jets production in single diffraction
(\ref{process}) offers another interesting example since factorization of hard and soft interactions in this case is less complicated.
In particular,  the fluctuation of a proton projectile into  a state with small transverse size,
which is the underlining mechanism for  (\ref{process}), suppresses secondary soft interactions that may fill the rapidity gap.
Thus one can get an access to the gluon distribution at small $x$ in a cleaner environment,
having no problems with gap
survival probability and factorization breaking that introduce major conceptual
theoretical uncertainties in the calculations of diffractive Higgs production.

Our approach to exclusive three-jet production derives from experience with coherent pion diffraction
dissociation into a pair of jets with large transverse momenta which was measured by
the E791 collaboration \cite{E791a,E791b}. The qualitative features of the E791 data have confirmed
 some earlier theoretical predictions \cite{KDR80,BBGG81,FMS93}:
a strong A-dependence which is  a signature for color transparency, and a $\sim 1/q_\perp^8$
dependence on the jet transverse momentum. These features suggest that
the relevant transverse size of the pion $r_\perp$ remains small,
of the order of the inverse transverse momenta of the jets $r_\perp  \sim 1/q_\perp$.

On a more quantitative level, we have shown \cite{Braun:2001ih,Braun:2002wu}  that
collinear factorization is violated in dijet production due
to pinching of singularities between soft gluon (and quark)
interactions in the initial and final state.
However, the nonfactorizable contribution is suppressed compared to the leading contribution by a logarithm of energy so that
 in  the double logarithmic approximation  $\ln q_\perp^2 \ln s/q_\perp^2$  collinear factorization is restored.
Moreover, to this accuracy  hard gluon exchange can be ``hidden'' in the  unintegrated gluon distribution
${\cal F}(x,q_\perp)$. Thus, in the true diffraction limit, for very large energies,
hard exclusive dijet production can be considered as a probe of the hard component of the pomeron.
The same interpretation  was suggested  earlier in \cite{NSS99} within the $k_t$ factorization framework
(see also \cite{Nikolaev:1994cd}).
The double logarithmic approximation turns out to be insufficient for the energy range of the E791 experiment,
but might be adequate for the LHC. In this Letter we present an estimate for the
cross section for the reaction (\ref{process})  based on the generalization of these ideas.

\begin{figure}[t]
\centerline{\epsfxsize6cm\epsffile{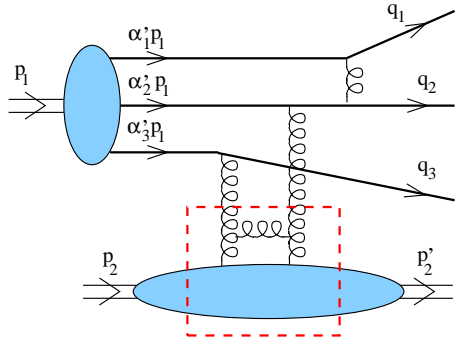}}
\caption{\small
Proton dissociation into three jets. The unintegrated gluon distribution
includes the hard gluon exchange as indicated by the dashed square.
}
\label{fig:figsum}
\end{figure}

\vskip0.2cm

{\large \bf 2.}~
At leading order the jets are formed by the three valence quarks of the proton,
see Fig.~1.
 We require that all three jets
have large transverse momenta which requires at least two hard gluon exchanges. One of them can be
effectively included in the high-momentum component of the unintegrated gluon density (the bottom blob)
as indicated schematically by the dashed square,
 but the second one has to be added explicitly since the  hard  pomeron only couples to two
of the three quarks of the proton.%
\footnote{Alternatively, one can consider double--pomeron exchange in the t-channel. This contribution is
 suppressed by a power of the jet transverse momentum so it is of higher twist.}
The necessity for an additional hard gluon exchange makes calculation of proton diffraction dissociation
 more difficult as compared to the meson case.

Our notation for the momenta is explained in Fig.~\ref{fig:figsum}.
We neglect power corrections in  transverse momenta of the jets and also proton and jet masses so that
$p_1^2=p_2^2=p_2^{\prime 2}=q_1^2=q_2^2=q_3^2=0.$
The jet momenta are decomposed in terms of momenta of the initial particles
\begin{eqnarray}
q_k=\alpha_k p_1+\beta_k p_2+q_{k\perp}\,, \qquad k=1,2,3
\end{eqnarray}
where
\begin{equation}
\vec{q}_{1\perp}+\vec{q}_{2\perp}+\vec{q}_{3\perp}=0\, , \qquad \alpha_1+\alpha_2+\alpha_3=1\,, \qquad
 \beta_k=\vec{q}_{k\perp }^{\,\, 2}/(\alpha_k s)\,,%
\end{equation}
The  three--jet invariant mass is given by
\begin{equation}
\label{zeta}
M^2=(q_1+q_2+q_3)^2=\frac{\vec q_{1\perp}^{\,\, 2}}{\alpha_1} +  \frac{\vec q_{2\perp}^{\,\,
2}}{\alpha_2}+ \frac{\vec q_{3\perp}^{\,\, 2}}{\alpha_3}\, ,\qquad
\zeta=\frac{M^2}{s} = \beta_1+\beta_2+\beta_3\,.
\end{equation}
where $s=(p_1+p_2)^2=2p_1\cdot p_2$ is the invariant energy.
Assuming that the  relevant  jet transverse momenta are of the order of 5~GeV, the
typical values of the $\zeta$ variable at LHC are in the range $\zeta\sim 10^{-6}\div 10^{-5}$.

At high energies, an amplitude is predominantly given by its discontinuity in
the s-channel which usually implies that the amplitude is almost purely imaginary.
In our case the situation is more complicated since in the physical region of (\ref{process})
the amplitude develops a cut in the variable $M^2$ as well, and it remains complex even after taking
the $s$-channel discontinuity.
Nevertheless, the $s$-channel discontinuity of the amplitude can be expressed,
in our approximation, in terms of the unintegrated gluon distribution
${\cal F}(x,k_\perp)$.

The relevant Feynman diagrams in leading order of perturbative QCD
are shown in Fig.~2.
\begin{figure}[t]
\centerline{
  \includegraphics[width=0.55\textwidth,angle=0]{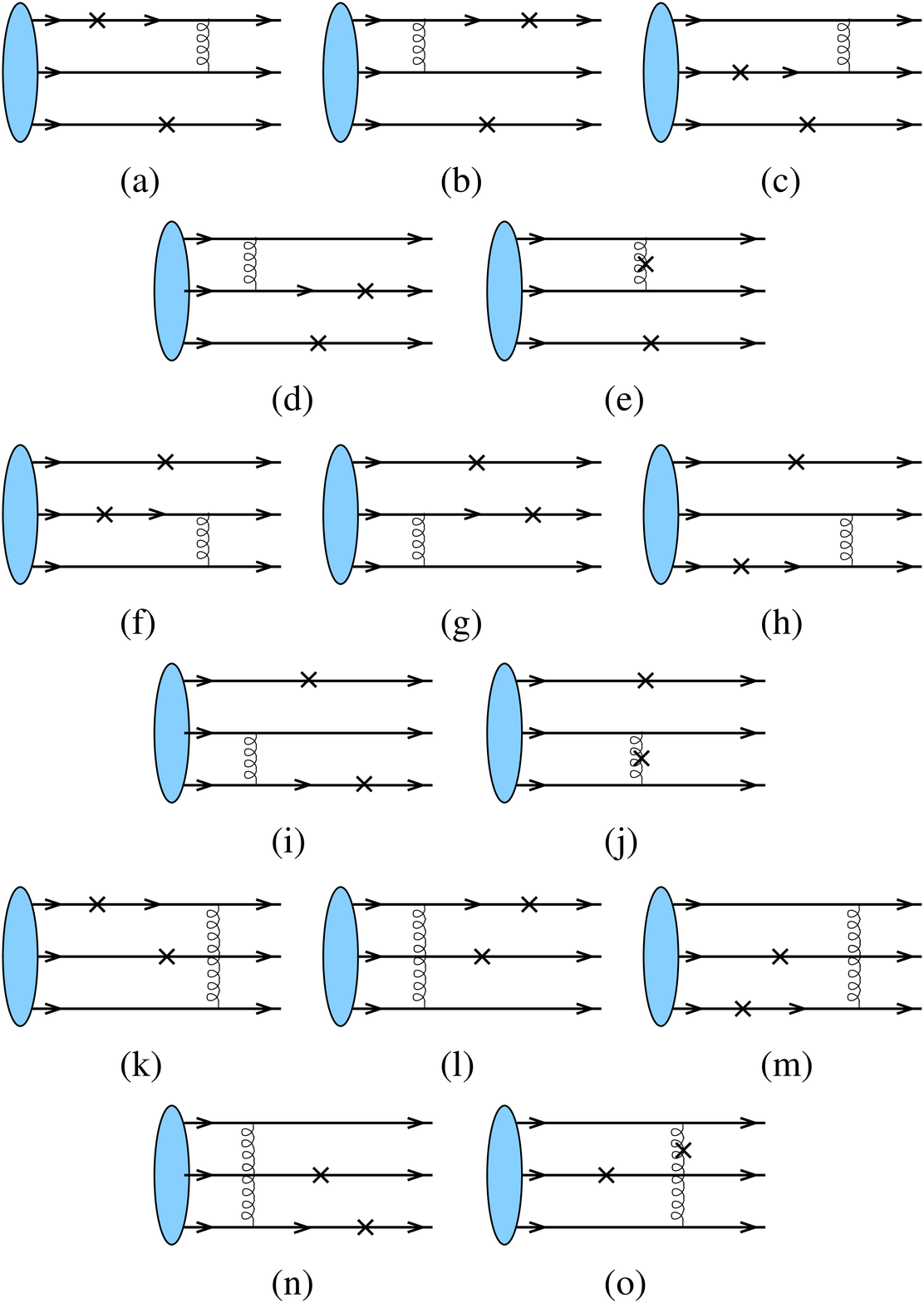}}
\caption{The leading-order contributions to proton disintegration into three jets 
within $k_t$ factorization.
The points where the t-channel gluons are attached to the quarks are shown by crosses.
}
\label{fig:G12}
\end{figure}
They can be  divided into three
groups which differ by the attachments of the $t$-channel gluons to the quark lines
(shown by crosses).
In diagrams (f)-(j) the hard gluon exchange takes place between quark $q_2$ and
$q_3$. Hence the transverse momentum of one of the $t$-channel gluons
 coincides with that of the quark $q_1$. As a consequence,
the contribution of this group of diagrams involves the unintegrated gluon distribution
at the same scale, ${\cal F}(\zeta ,q_{1\perp})$.
In diagrams (k)-(o) the hard gluon exchange connects the quarks $q_1$ and
$q_3$, the transverse momentum of the $t$-channel gluon coincides with the momentum of
the quark $q_2$ and, therefore, the unintegrated gluon distribution
enters at this scale, ${\cal F}(\zeta ,q_{2\perp})$.
Similarly, the contribution of  diagrams (a)-(e) is proportional to ${\cal F}(\zeta,q_{3\perp})$.

Accordingly, we have three different
contributions to the amplitude:
\begin{eqnarray}
\label{ampl}
 \mathcal{M} & = & -i\,   2^7 \pi^5\, s \, \alpha_s^2
\left[\frac{e^{ijk}\left(\frac{1+N}{N}\right)^2}{4N!(N^2-1)}\right]\, \int
D\alpha^\prime\,
 \\
&&\hspace*{-1cm}\times \left(
\mathcal{L}_{f \div j}\frac{\delta(\alpha_1-\alpha_1^\prime)}{q_{1\perp}^4}{\mathcal F}(\zeta,q_{1\perp})+
\mathcal{L}_{k\div o}\frac{\delta(\alpha_2-\alpha_2^\prime)}{q_{2\perp}^4}{\mathcal F}(\zeta,q_{2\perp})+
\mathcal{L}_{a\div e}\frac{\delta(\alpha_3-\alpha_3^\prime)}{q_{3\perp}^4}{\mathcal F}(\zeta,q_{3\perp})\right) \, ,
 \nonumber
\end{eqnarray}
where $\int D\alpha' = \int_0^1 d\alpha'_1 d\alpha'_2 d\alpha'_3 \delta(1-\sum \alpha'_i)$ corresponds to the integration over the quark momentum fractions in the incident proton,
$e^{ijk}$ describes the color state of the final quarks, $N=3$ is the number of colors.
The dimensionless quantities $\mathcal{L}_i$ are expressed in terms of different
Dirac structures where for convenience we introduce a "positron-like"
Dirac spinor $v$,
\begin{equation}
\left(\bar u(q_2)\right)^T=C \, v(q_2)\, , \quad
\left(\bar u(q_2)\gamma_\mu\right)^T=-C \gamma_\mu \, v(q_2) \,.
\end{equation}
Here $C$ is the charge-conjugation matrix.
The diagrams with the 3-gluon vertex do not contribute due to vanishing color factors.
Therefore we need to calculate in total 12 nontrivial diagrams (4 for each group).

The calculation is straightforward, though rather
tedious. Here we present the final results only:
\begin{eqnarray}
\label{L1}
\mathcal{L}_{f\div j}&=&\left[ {\cal V} \, \bar u(q_1)\!\!\not\!p_1 v(q_2)\,
\bar u(q_3)\!\!\not\!p_2 \gamma_5 N(p_1) -
{\cal A}\, \bar u(q_1)\!\!\not\! p_1\gamma_5 v(q_2)\, \bar u(q_3)\!\!\not\!p_2
N(p_1) \right]
\nonumber\\
&&\times
\left(\frac{\alpha_1(\alpha_3+\alpha_3^\prime)}{[-\alpha_2^\prime\alpha_3\beta_2 s^3]
[\alpha_3^\prime(\beta_2+\beta_3)-\alpha_1\beta_1+i\epsilon]}+
\frac{-\alpha_1(\alpha_3+\alpha_3^\prime)}{[-\alpha_3^\prime\alpha_3\beta_3 s^3]
[\alpha_2^\prime(\beta_2+\beta_3)-\alpha_1\beta_1+i\epsilon]}
\right)
\nonumber \\
&&+
\left[ {\cal V} \, \bar u(q_1)\!\!\not\!p_2 v(q_2)\, \bar u(q_3)\!\!\not\!p_2 \gamma_5 N(p_1) -
{\cal A}\, \bar u(q_1)\!\!\not\! p_2\gamma_5 v(q_2)\, \bar u(q_3)\!\!\not\!p_2
N(p_1) \right]
\nonumber \\
&&\times
\left(\frac{\alpha_2^\prime\beta_2-\alpha_1\beta_1}{[-\alpha_3^\prime\alpha_3\beta_3 s^3]
[\alpha_2^\prime(\beta_2+\beta_3)-\alpha_1\beta_1+i\epsilon]}
+\frac{\alpha_1\beta_1+\alpha_3\beta_3+(\alpha_3-\alpha_2^\prime)\beta_2}
{[-\alpha_2^\prime\alpha_3\beta_2 s^3]
[\alpha_3^\prime(\beta_2+\beta_3)-\alpha_1\beta_1+i\epsilon]}
\right.\nonumber\\ &&{}\hspace*{0.5cm}\left.
+\frac{\alpha_2^\prime\beta_2+\alpha_3^\prime\beta_3}
{\alpha_2^\prime\alpha_3^\prime\beta_2\beta_3(\alpha_2+\alpha_3)s^3}
\right)
\nonumber \\
&&+
\left[ {\cal V} \, \bar u(q_1)\!\!\not\!p_1 \gamma_5 v(q_2)\, \bar u(q_3)\!\!\not\!p_2  N(p_1) -
{\cal A}\, \bar u(q_1)\!\!\not\! p_1 v(q_2)\, \bar u(q_3)\!\!\not\!p_2
\gamma_5 N(p_1) \right]
\nonumber \\
&&\times
\left(\frac{-\alpha_1(\alpha_2+\alpha_2^\prime)}{[-\alpha_2^\prime\alpha_3\beta_2 s^3]
[\alpha_3^\prime(\beta_2+\beta_3)-\alpha_1\beta_1+i\epsilon]}+
\frac{\alpha_1(\alpha_2+\alpha_2^\prime)}{[-\alpha_3^\prime\alpha_3\beta_3 s^3]
[\alpha_2^\prime(\beta_2+\beta_3)-\alpha_1\beta_1+i\epsilon]}
\right)
\nonumber \\
&&{}+
\left[ {\cal V} \, \bar u(q_1)\!\!\not\!p_2 \gamma_5 v(q_2)\, \bar u(q_3)\!\!\not\!p_2  N(p_1) -
{\cal A}\, \bar u(q_1)\!\!\not\! p_2 v(q_2)\, \bar u(q_3)\!\!\not\!p_2
\gamma_5 N(p_1) \right]
\nonumber \\
&&\times
\left(\frac{\alpha_2^\prime\beta_2+\alpha_1\beta_1}{[-\alpha_3^\prime\alpha_3\beta_3 s^3]
[\alpha_2^\prime(\beta_2+\beta_3)-\alpha_1\beta_1+i\epsilon]}
+\frac{-\alpha_1\beta_1+\alpha_3\beta_3+(\alpha_3-\alpha_2^\prime)\beta_2}
{[-\alpha_2^\prime\alpha_3\beta_2 s^3]
[\alpha_3^\prime(\beta_2+\beta_3)-\alpha_1\beta_1+i\epsilon]}
\right.\nonumber\\ &&{}\hspace*{0.5cm}\left.
+\frac{\alpha_2^\prime\beta_2+\alpha_3^\prime\beta_3}
{\alpha_2^\prime\alpha_3^\prime\beta_2\beta_3(\alpha_2+\alpha_3)s^3}
\right)
\nonumber \\
&&{}+
{\cal T}\left[ \bar u(q_1) v(q_2)\, \bar u(q_3)\gamma_5 N(p_1) -
 \bar u(q_1)\gamma_5 v(q_2)\, \bar u(q_3)
N(p_1) \right]
\nonumber \\
&&\times
\left(\frac{\alpha_3^\prime-\alpha_1}{[-\alpha_2^\prime\beta_2 s^2]
[\alpha_3^\prime(\beta_2+\beta_3)-\alpha_1\beta_1+i\epsilon]}
+\frac{(\alpha_2-\alpha_1)\beta_2+(\alpha_2-\alpha_3^\prime)\beta_3}{[-\alpha_3^\prime\beta_3^2 s^2]
[\alpha_2^\prime(\beta_2+\beta_3)-\alpha_1\beta_1+i\epsilon]}
\right.\nonumber\\ &&{}\hspace*{0.5cm}\left.
+\frac{(\alpha_2-\alpha_1)(\alpha_2^\prime\beta_2+\alpha_3^\prime\beta_3)}
{\alpha_2^\prime\alpha_3^\prime\beta_2\beta_3^2(\alpha_2+\alpha_3)s^2}
\right)
\nonumber \\
&&{}+
{\cal T}\left[ \bar u(q_1)\left(i p_{1\mu}p_{2\nu}\sigma_{\mu\nu}\right) v(q_2)\,
\bar u(q_3)\gamma_5 N(p_1) -
 \bar u(q_1)\left(i p_{1\mu}p_{2\nu}\sigma_{\mu\nu}\right)\gamma_5 v(q_2)\, \bar u(q_3)
N(p_1) \right]
\nonumber \\
&&\times
\left(\frac{2(1-\alpha_2^\prime)}{[-\alpha_2^\prime\beta_2 s^3]
[\alpha_3^\prime(\beta_2+\beta_3)-\alpha_1\beta_1+i\epsilon]}
+ \frac{2((\alpha_1+\alpha_2)\beta_2+(\alpha_2-\alpha_3^\prime)\beta_3)}
{[-\alpha_3^\prime\beta_3^2 s^3]
[\alpha_2^\prime(\beta_2+\beta_3)-\alpha_1\beta_1+i\epsilon]}
\right.\nonumber\\ &&{}\hspace*{0.5cm}\left.
+\frac{2(\alpha_1+\alpha_2)(\alpha_2^\prime\beta_2+\alpha_3^\prime\beta_3)}
{\alpha_2^\prime\alpha_3^\prime\beta_2\beta_3^2(\alpha_2+\alpha_3)s^3}
\right),
\end{eqnarray}
\begin{eqnarray}
\label{L3}
\mathcal{L}_{a\div e}&=&
\left[ {\cal V} \, \bar u(q_1)\!\!\not\!p_1 v(q_2)\, \bar u(q_3)\!\!\not\!p_2 \gamma_5 N(p_1) -
{\cal A}\, \bar u(q_1)\!\!\not\! p_1\gamma_5 v(q_2)\, \bar u(q_3)\!\!\not\!p_2
N(p_1) \right]
\nonumber\\
&&\times
\left(\frac{2(\alpha_2^\prime-\alpha_1)}{[-\alpha_2^\prime\beta_2 s^3]
[\alpha_1^\prime(\beta_1+\beta_2)-\alpha_3\beta_3+i\epsilon]}+
\frac{2(\alpha_1^\prime-\alpha_2)}{[-\alpha_1^\prime\beta_1 s^3]
[\alpha_2^\prime(\beta_1+\beta_2)-\alpha_3\beta_3+i\epsilon]}
\right)
\nonumber \\
&&{}+
\left[ {\cal V} \, \bar u(q_1)\!\!\not\!p_2 v(q_2)\, \bar u(q_3)\!\!\not\!p_2 \gamma_5 N(p_1) -
{\cal A}\, \bar u(q_1)\!\!\not\! p_2\gamma_5 v(q_2)\, \bar u(q_3)\!\!\not\!p_2
N(p_1) \right]
\nonumber \\
&&\times
\left(\frac{2\beta_1}{[-\alpha_2^\prime\beta_2 s^3]
[\alpha_1^\prime(\beta_1+\beta_2)-\alpha_3\beta_3+i\epsilon]}
+\frac{2\beta_2}{[-\alpha_1^\prime\beta_1 s^3]
[\alpha_2^\prime(\beta_1+\beta_2)-\alpha_3\beta_3+i\epsilon]}
\right.\nonumber\\ &&{}\hspace*{0.5cm}\left.
+\frac{2(\alpha_1^\prime\beta_1+\alpha_2^\prime\beta_2)}
{\alpha_1^\prime\alpha_2^\prime\beta_1\beta_2(\alpha_1+\alpha_2)s^3}
\right)
\nonumber \\
&&{}+
{\cal T}\left[ \bar u(q_1) v(q_2)\, \bar u(q_3)\gamma_5 N(p_1) -
 \bar u(q_1)\gamma_5 v(q_2)\, \bar u(q_3)
N(p_1) \right]
\nonumber \\
&&\times
\left(\frac{-2\alpha_3}{[-\alpha_2^\prime\beta_2 s^2]
[\alpha_1^\prime(\beta_1+\beta_2)-\alpha_3\beta_3+i\epsilon]}
+\frac{2\alpha_3}{[-\alpha_1^\prime\beta_1 s^2]
[\alpha_2^\prime(\beta_1+\beta_2)-\alpha_3\beta_3+i\epsilon]}
\right).
\end{eqnarray}
The functions $\mathcal{A}(\alpha'_1, \alpha'_2, \alpha'_3)$,  $\mathcal{V}(\alpha'_1, \alpha'_2, \alpha'_3)$ and
$\mathcal{T}(\alpha'_1, \alpha'_2, \alpha'_3)$ are the leading-twist light-cone nucleon distribution
amplitudes defined as in \cite{Chernyak:1983ej}
\begin{eqnarray}
\lefteqn{
\langle 0|\epsilon^{ijk}u^i_{\alpha}(a_1 z)u^j_{\beta}(a_2 z)d^k_{\gamma}(a_3
z)|N(p_1)\rangle \, = }
\nonumber \\
&=&V \left(\not p_1C\right)_{\alpha\beta}\left(\gamma_5
N(p_1)\right)_\gamma +
A\left(\not p_1\gamma_5 C\right)_{\alpha\beta}\left(
N(p_1)\right)_\gamma +
T\left(i \sigma_{\mu \nu}p^\nu_1 C\right)_{\alpha\beta}\left(
\gamma^\mu\gamma_5N(p_1)\right)_\gamma \, ,
\label{DA}
\end{eqnarray}
where $z^2=0$ and $\sigma_{\mu\nu}=(i/2)[\gamma_{\mu},\gamma_{\nu}]$.
Each invariant amplitude $V, A, T$ depends on the scalar products $a_i p_1\cdot z$
and can be represented as the Fourier transform of the corresponding distribution
amplitude, e.g.
\begin{equation}
V(a_i p_1 z)=\int D\alpha^\prime e^{-ip_1\cdot z \sum_i\alpha^\prime_i
a_i}{\mathcal V}(\alpha^\prime_i)\,.
\end{equation}

{}Finally, $\mathcal{L}_{k-o}$ is given by the expression similar to $\mathcal{L}_{f-j}$ in Eq.~(\ref{L1})
with the replacements $\alpha_1\leftrightarrow \alpha_2$, $\alpha'_1\leftrightarrow \alpha'_2$ and
$\beta_1\leftrightarrow \beta_2$ everywhere except for the arguments of the distribution
amplitudes.

One remark is in order. Another possible mechanism for the exclusive proton disintegration into  three jets could be
the exchange of three hard gluons  in the $t$- channel.
 Such a contribution could involve two different color structures, proportional to $f^{abc}$ and $d^{abc}$,
which correspond to different  $C$-parity in the $t-$ channel.
We found by explicit calculation that the color factor $\sim f^{abc}$ corresponding to the $C$-parity-even exchange
vanishes.
The $C$- parity odd contribution  $\sim d^{abc}$  is related to odderon exchange and presumably small.

\vskip0.2cm

{\large \bf 3.}~The differential cross section can be written as
\begin{equation}
d\sigma =\frac{|\mathcal{M}|^2}{2^5(2\pi)^8 s^2}
\frac{d\alpha_1 d\alpha_2 d\alpha_3 \delta(1-\alpha_1-\alpha_2-\alpha_3)}{\alpha_1\alpha_2\alpha_3}
d^2\vec q_{1}d^2\vec q_{2} dt d\phi_t
\end{equation}
where $t=(p_2-p_2^\prime)^2$ is the Mandelstam $t$ variable of the
$pp$ scattering and $\phi_t$ is
the azimuthal angle of the final state proton. In our kinematics, for large transverse momenta
of the jets and small $t$, one can neglect effects of azimuthal correlations between the jets and
the final proton. Hence  $d\phi_t$  integration is trivial and gives a factor $2\pi$.
For the $t$ dependence we assume a simple
exponential form, $d\sigma/dt \sim e^{b t}$,
and use $b\sim 4\div 5 \, \rm{GeV}^2$ for the slope parameter which is a typical value which describes HERA data for hard
exclusive processes: DVCS and vector meson electroproduction at large $Q^2$.
Thus, the integration over the proton recoil  variables gives a factor
$\int dt d\phi_t \to \frac{2\pi}{b}$.

Since our calculation is only done to double logarithmic accuracy,
 we use the simplest model for the unintegrated gluon distribution as given by the
logarithmic derivative of the usual gluon parton distribution $x g (x,Q^2)$
\begin{equation}
\label{unintegrated}
{\mathcal F}(x,q_{\perp}^2)={\partial \over \partial \ln q_\perp^2}\, x\,
g(x,q_\perp^2) \,.
\end{equation}
The numerical estimates presented below are obtained using the CTEQ6L leading-order
gluon distribution as provided by \cite{Pumplin:2002vw}. We also used the simplest, asymptotic
model for the nucleon distribution amplitude
\begin{equation}
 \label{asym}
{\mathcal V}(\alpha'_i)\,=\,{\mathcal T}(\alpha'_i)\,=\,120 f_N \alpha'_1\alpha'_2\alpha'_3 \,, \qquad {\mathcal A}(\alpha'_i)\,=\,0\,.
\end{equation}
The normalization parameter $f_N$ is scale-dependent. To leading-logarithmic accuracy
\begin{equation}
f_N(\mu)=f_N(\mu_0)\left(\frac{\alpha_s(\mu)}{\alpha_s(\mu_0)}\right)^{\frac{2}{3\beta_0}} \,,
\end{equation}
where $\beta_0 = 11/3 N -2/3 n_f$. The existing QCD sum rule estimates
\begin{eqnarray}
m_N f_N(\mu = 1~\mbox{GeV}) &=& (5.0\pm 0.3)\times 10^{-3}\,\rm{GeV}^3\quad \cite{Chernyak:1987nu}\,,
\nonumber\\
m_N f_N(\mu = 1~\mbox{GeV}) &=& (5.1\pm 0.4)\times 10^{-3}\,\rm{GeV}^3\quad \cite{King:1986wi}\,
\end{eqnarray}
are somewhat larger compared  with a very recent $n_f=2$ (unquenched) lattice calculation
\begin{equation}
 f_N(\mu = 2~\mbox{GeV}) = (3.14\pm 0.09)\times 10^{-3}\,\rm{GeV}^2\quad \cite{Gockeler:2008xv}\,.
\end{equation}
The given number corresponds to the lattice spacing $a\simeq 0.067$~fm; a continuum extrapolation
was not attempted. For definiteness we use the value $f_N=5.0 \times 10^{-3}\,\rm{GeV}^2$ at 1~GeV as input,
and evolve it to the relevant scale.

The integration over the phase space of the three jets was done numerically, restricting the
longitudinal momentum fractions to the region%
\footnote{We use this rather conservative cut condition to assure a clear three--jet event selection.}
\begin{equation}
0.1 \leq \alpha_1,\alpha_2,\alpha_3 \leq 0.8
\end{equation}
and requiring that the transverse momentum of {\it each} jet is larger than a given
value $q_0 = q_{\perp, {\rm min }}$. For the value $q_0 = 5$~GeV
we obtain for the integrated three-jet cross section at the LHC energies
\begin{equation}
 \sigma^{\rm LHC}_{3-jets} = 4\,\mbox{pb}\,\cdot \left(\frac{f_N(q_0)}{4.7\cdot 10^{-3}\mbox{\small GeV}^2}\right)^2
 \left(\frac{\alpha_s(q_0)}{0.21}\right)^4 \left(\frac{5 \,\mbox{\small GeV}}{q_0}\right)^9.
\end{equation}
Assuming the integrated luminosity for the first LHC runs in the range $100$~pb$^{-1}$ to  $1$~fb$^{-1}$
an observation of this process at LHC seems to be feasible.
Note that the effective power $\sigma \sim1/q^9_0$ (fitted in the $q_0=3\div 8$~GeV range) is somewhat stronger
than the naive power counting prediction $\sigma\sim 1/q_0^8$. This effect is due to the strong $\zeta$ dependence
of the unintegrated gluon distribution: larger values of $q_0$ imply larger invariant masses $M^2$ of the
three-jet system (\ref{zeta}) and consequently  larger $\zeta=M^2/s$.
The sizeable cross section for $q_0=5$~GeV is in fact an implication of the expected rise of
the LO gluon distribution more than two times as $\zeta$ is decreasing by roughly a factor of $50$
when going from  Tevatron to LHC.
The existing parameterizations of the LO gluon distribution at $\zeta\sim 10^{-6}$ differ from each other by $\sim 30\%$.
The unintegrated gluon distribution (\ref{unintegrated}) enters as a square in the prediction for the cross section,
therefore, the study of exclusive three-jet events at LHC may provide a valuable constraint
for the gluon distribution at small momentum fractions.

A comparison of the three-jet exclusive production at LHC and the Tevatron can be especially illuminating
in this respect since other uncertainties do not have significant impact on  the energy dependence.
For Tevatron kinematics, assuming the value $q_{\perp \rm{min}}= 3$~GeV, our estimate for the cross section (fitted in the range $q_0=2\div 4.5$~GeV) is
\begin{equation}
\label{tevatron}
 \sigma^{\rm Tevatron}_{3-jets} = 50\,\mbox{pb}\,\cdot
 \left(\frac{f_N(q_0)}{4.7\cdot 10^{-3}\mbox{\small GeV}^2}\right)^2
 \left(\frac{\alpha_s(q_0)}{0.255}\right)^4 \left(\frac{3\,\mbox{\small GeV}}{q_0}\right)^9.
\end{equation}
Note that in this case $M^2\sim 100\, \rm{GeV}^2$ and $\zeta\sim 10^{-5}\div 10^{-4}$ where
the gluon distribution is much better known: The typical difference between existing parameterizations
is of order $\sim 10\%$.

\begin{figure}[t]
\centerline{
  \includegraphics[width=0.55\textwidth,angle=0]{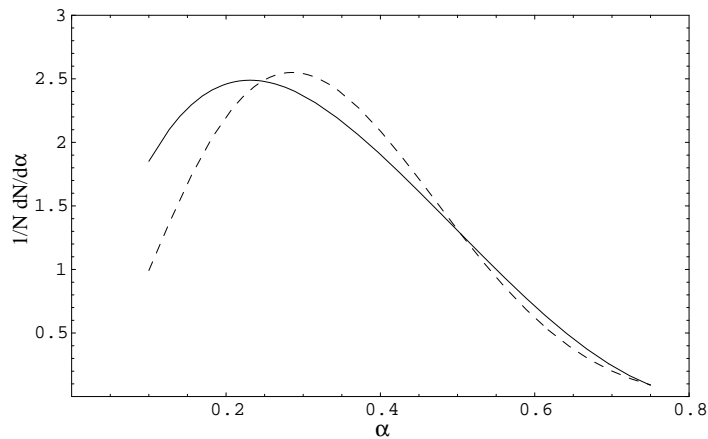}}
\caption{The normalized longitudinal momentum fraction distribution of the jet (solid curve).
For comparison, the dependence corresponding to Eq.~(\ref{longnaive}) is shown by dashes.
}
\label{fig:long}
\end{figure}

In the most naive approximation, the longitudinal momentum fraction distribution 
of the jets is expected to follow that of the valence quarks in the proton:
The momentum fraction distribution of a jet, arbitrary chosen in each event, 
is proportional to the proton distribution amplitude squared 
\begin{equation}
\label{longnaive}
\frac{d\sigma}{d\alpha}
 \sim \int D\alpha' \delta(\alpha-\alpha'_1) |\phi_N(\alpha')|^2\, ,
\end{equation}
 where $\phi_N = \mathcal{V}-\mathcal{A}$ \cite{Chernyak:1983ej}.
In reality, a hard gluon exchange leads to a certain redistribution of the longitudinal momenta so that the
resulting $\alpha$-dependence  is more complex, see e.g. \cite{Braun:2001ih,Braun:2002wu}.
To illustrate this effect, in Fig.~\ref{fig:long} we compare our calculated normalized jet momentum fraction distribution
(averaged over quark flavors) to the dependence in (\ref{longnaive}): The two distributions are similar, but the
one resulting from the QCD calculation is shifted towards lower momentum fractions compared to the simple
dependence in Eq.~(\ref{longnaive}).
Note that measurement of the valence quark momentum fraction distribution in a pion presented the main motivation
for the E791 experiment \cite{E791a,E791b} which, in turn, triggered detailed  studies of such reactions in pQCD.

\vskip0.2cm

{\large \bf 4.}~To summarize, in this Letter we have studied the exclusive diffractive dissociation of a proton
into three jets with large transverse momenta in the double-logarithmic approximation of perturbative QCD.
This process is interesting in the broader context of diffractive processes at LHC, which will be
studied using forward detectors, and in particular can be used to constrain the gluon distribution
at very small values of Bjorken $x$.
According to our estimates, an observation of such processes in the early runs at LHC is
feasible for jet transverse momenta of the order of $5$~GeV.

\vskip0.2cm

{\bf Acknowledgments }
We gratefully acknowledge participation of  S.~Gottwald in this project on early stages.
 D.~I.\ thanks the Institute for Theoretical Physics at the University of Regensburg
for hospitality. His work was  supported in part by the
BMBF-GSI, project number 06RY258, and  grants RFBR-06-02-16064-a and NSh 1027.2008.2.



\begin{thebibliography}{99}

\bibitem{Joram:2006kf}
  C.~Joram, A.~Braem and H.~Stenzel,\  ATL-LUM-PUB-2007-002.

\bibitem{Royon:2007ah}
  C.~Royon  [RP220 Collaboration],
  arXiv:0706.1796 [physics.ins-det].

\bibitem{Rouby:2008aj}
  X.~Rouby  [CMS Collaboration],
  arXiv:0805.4406 [hep-ex].

\bibitem{Albrow:2008pn}
  M.~G.~Albrow {\it et al.}  [FP420 R\&D Collaboration],
  arXiv:0806.0302 [hep-ex].

\bibitem{d'Enterria:2008is}
  D.~d'Enterria,
  arXiv:0806.0883 [hep-ex].


\bibitem{Frankfurt:2002jq}
  L.~Frankfurt and M.~Strikman,
  arXiv:hep-ph/0210087;
  Phys.\ Rev.\  D {\bf 67} (2003) 017502.

\bibitem{Aaltonen:2007hs}
  T.~Aaltonen {\it et al.}  [CDF Run II Collaboration],
  Phys.\ Rev.\  D {\bf 77} (2008) 052004.

\bibitem{Khoze:2008cx}
  V.~A.~Khoze, A.~D.~Martin and M.~G.~Ryskin,
  arXiv:0802.0177 [hep-ph].

\bibitem{E791a}
E.~M.~Aitala {\it et al.}  [E791 Collaboration],
Phys.\ Rev.\ Lett.\  {\bf 86} (2001) 4768.

\bibitem{E791b}
E.~M.~Aitala {\it et al.}  [E791 Collaboration],
Phys.\ Rev.\ Lett.\  {\bf 86} (2001) 4773.


\bibitem{KDR80}
S.~F.~King, A.~Donnachie and J.~Randa,
Nucl.\ Phys.\ B {\bf 167} (1980) 98;\\
J.~Randa,
Phys.\ Rev.\ D {\bf 22} (1980) 1583.

\bibitem{BBGG81}
G.~Bertsch, S.~J.~Brodsky, A.~S.~Goldhaber and J.~F.~Gunion,
Phys.\ Rev.\ Lett.\ {\bf 47} (1981) 297.

\bibitem{FMS93}
L.~Frankfurt, G.~A.~Miller and M.~Strikman,
Phys.\ Lett.\ B {\bf 304} (1993) 1.

\bibitem{Braun:2001ih}
  V.~M.~Braun, D.~Yu.~Ivanov, A.~Sch{\"a}fer and L.~Szymanowski,
  Phys.\ Lett.\  B {\bf 509} (2001) 43.


\bibitem{Braun:2002wu}
  V.~M.~Braun, D.~Yu.~Ivanov, A.~Sch{\"a}fer and L.~Szymanowski,
  Nucl.\ Phys.\  B {\bf 638} (2002) 111.

\bibitem{NSS99}
  N.~N.~Nikolaev, W.~Sch{\"a}fer and G.~Schwiete,
  Phys.\ Rev.\  D {\bf 63} (2001) 014020.

\bibitem{Nikolaev:1994cd}
  N.~N.~Nikolaev and B.~G.~Zakharov,
  Phys.\ Lett.\  B {\bf 332} (1994) 177.

\bibitem{Chernyak:1983ej}
  V.~L.~Chernyak and A.~R.~Zhitnitsky,
  Phys.\ Rept.\  {\bf 112} (1984) 173.

\bibitem{Pumplin:2002vw}
  J.~Pumplin, D.~R.~Stump, J.~Huston, H.~L.~Lai, P.~Nadolsky and W.~K.~Tung,
  JHEP {\bf 0207} (2002) 012;
{\it http://user.pa.msu.edu/wkt/cteq/cteq6/cteq6pdf.html}

\bibitem{Chernyak:1987nu}
  V.~L.~Chernyak and I.~R.~Zhitnitsky,
  Nucl.\ Phys.\  B {\bf 246} (1984) 52;
 V.~L.~Chernyak, A.~A.~Ogloblin and I.~R.~Zhitnitsky,
  Z.\ Phys.\  C {\bf 42} (1989) 569.

\bibitem{King:1986wi}
  I.~D.~King and C.~T.~Sachrajda,
  Nucl.\ Phys.\  B {\bf 279} (1987) 785.

\bibitem{Gockeler:2008xv}
  M.~G{\"o}ckeler {\it et al.},
  arXiv:0804.1877 [hep-lat].

\end{thebibliography}
\end{document}